\documentclass[a4paper,fleqn,usenatbib]{mnras}

\usepackage{mathptmx}
\usepackage[T1]{fontenc}
\usepackage{ae,aecompl}
\usepackage{soul}

\usepackage{graphicx}	% Including figure files
\usepackage{amsmath}	% Advanced maths commands
\usepackage{amssymb}	% Extra maths symbols

\newcommand\etal{{et al. }}
\newcommand\angstrom{\text{\AA}}
\providecommand{\e}[1]{\ensuremath{\times 10^{#1}}}
\title[Search for J0337 Pulses]{Search for Optical Pulsations in PSR J0337+1715}

\author[M. J. Strader et al.]{
M. J. Strader,$^{1}$\thanks{E-mail: mstrader@physics.ucsb.edu}
A. M. Archibald,$^{2}$
S. R. Meeker,$^{1}$
P. Szypryt,$^{1}$
A. B. Walter,$^{1}$
\newauthor
J.C. van Eyken,$^{3}$
G. Ulbricht,$^{1}$
C. Stoughton,$^{4}$
B. Bumble,$^{5}$
D. L. Kaplan,$^{6}$
\newauthor
and B. A. Mazin$^{1}$
\\
$^{1}$Department of Physics, University of California, Santa Barbara, CA 93106, USA\\
$^{2}$ASTRON, the Netherlands Institute for Radio Astronomy, Postbus 2, 7990 AA, Dwingeloo, The Netherlands\\
$^{3}$NASA Exoplanet Science Institute, California Institute of Technology, 770 South Wilson Avenue, M/S 100-22, Pasadena, CA 91125, USA\\
$^{4}$Fermilab Center for Particle Astrophysics, Batavia, IL 60510, USA\\
$^{5}$NASA Jet Propulsion Laboratory, 4800 Oak Grove Drive, Pasadena, CA 91109, USA\\
$^{6}$Department of Physics, University of Wisconsin-Milwaukee, 1900 E. Kenwood Boulevard, Milwaukee, WI 53211, USA
}

\pubyear{2015}

\begin{document}
\label{firstpage}
\pagerange{\pageref{firstpage}--\pageref{lastpage}}
\maketitle

\begin{abstract}
    We report on a search for optical pulsations from PSR J0337+1715 at its observed radio pulse period. PSR J0337+1715 is a millisecond pulsar (2.7\,ms spin period) in a triple hierarchical system with two white dwarfs, and has a known optical counterpart with g-band magnitude 18.  The observations were done with the Array Camera for Optical to Near-IR Spectrophotometry at the 200" Hale telescope at Palomar Observatory.  No significant pulsations were found in the range 4000-11000 \angstrom, and we can limit pulsed emission in g-band to be fainter than 25\,mag.
\end{abstract}

\begin{keywords}
pulsars: general -- pulsars: individual (PSR J0337+1715)  --  stars: neutron
\end{keywords}

%%%%%%%%%%%%%%%%%%%%%%%%%%%%%%%%%%%%%%%%%%%%%%%%%%

%%%%%%%%%%%%%%%%% BODY OF PAPER %%%%%%%%%%%%%%%%%%

\section{Introduction}

Millisecond pulsars (MSPs) are distinguished from classical rotation-powered pulsars (RPP) in their characteristic parameters.  MSPs have shorter periods, smaller period derivatives, and weaker magnetic fields.  They are thought to be classical pulsars that have been spun up by accretion from a binary companion \citep{Alpar1982,Bhattacharya1991}.  Many of the known MSPs reside in binary systems, generally with a low mass white dwarf companion.  

The optical emission properties of MSPs are completely unknown, though other types of pulsars are seen in the optical band. Optical emission can be thermal, originating from the neutron star surface, or non-thermal, originating from particles accelerated in the magnetosphere or from X-ray reprocessing in a circumstellar debris disk.  There are about thirty optical, ultraviolet (UV), or near-infrared identifications of isolated neutron stars.  Of the RPPs, pulse timing in these bands has been achieved for only five objects \citep{Mignani2011a}.   

White dwarf companions dominate any optical emission in a binary system, so most searches for MSP optical emission have focused on deep imaging of isolated MSPs.  These attempts at deep imaging have not resulted in any definitive optical detections \citep{Sutaria2003,Mignani2003,Koptsevich2003,Mignani2004}, although \citet{Sutaria2003} did find a possible $m_{V}\approx 25$\,mag counterpart to PSR J1024-0719.  Also, \citet{Kargaltsev2004} were able to observe UV emission from the neutron star in the spectrum of the binary MSP J0437-4715.  This UV emission was ascribed to thermal emission.

Another approach to find optical counterparts of MSPs is to search for pulses, using a known rotational period from observations in radio, X-rays, or $\gamma$-rays.  This can be done for pulsars in binary systems as well as for isolated pulsars.  Optical pulses have not yet been found in any MSP.  Optical timing is a challenge for millisecond pulsars due to their intrinsic faintness in the optical band and high pulsation frequencies.  To perform this measurement, detectors need high time resolution and high quantum efficiency.  There are just a few attempts in the literature to search for pulsed optical emission from MSPs, in particular, PSR B1937+21 and several MSPs in compact globular clusters, all with negative results \citep{Manchester1984,Middleditch1988}.

\subsection{PSR J0337+1715}
PSR J0337+1715 (hereafter J0337) is an MSP that resides in a hierarchical triple system with two white dwarves orbiting it \citep{Ransom2014}.  The white dwarves have masses 0.2 $M_{\odot}$ (inner) and 0.4 $M_{\odot}$ (outer), with orbital periods 1.6 days and 327 days, respectively.  Spectroscopy of the inner white dwarf has revealed that it has a surprisingly high temperature of 15800 K \citep{Kaplan2014}.  Discovery of the system has generated interest in using the system to test the strong equivalence principle of general relativity and to study the secular evolution of multi-body systems \citep{Ransom2014}.

For our purposes, the 18th magnitude inner white dwarf makes J0337 a convenient target for observing with our instrument, the ARray Camera for Optical to Near-IR Spectrophotometry (ARCONS; \citealt{Mazin2013}).  The white dwarf allows us to see the system in one second integration images, so that we can check in real time that it is placed on the science array away from areas of dead pixels.  The downside of course, is that the light from the white dwarf adds to the background noise, limiting the signal-to-noise ratio (SNR) that we could reach in our allotted time.

\subsection{ARCONS}

ARCONS utilizes a 2024 pixel array of Microwave Kinetic Inductance Detectors (MKIDs). These devices detect individual photons with high time resolution (2\,$\mu$s) and with energy resolution ($R=E/\Delta E \sim 8$ at 4000 \angstrom ).  
ARCONS has a passband of 4000-11000 \angstrom.  The instrument was deployed at the Coud\'{e} focus of the 200-inch Hale Telescope at Palomar Observatory. ARCONS has previously been used in optical observations of the Crab pulsar \citep{Strader2013}.

\section{Observations and Analysis}

We observed J0337 from the 200-inch Hale Telescope for a total of 6.7 hours on the nights of September 24,  September 25, and October 21, 2014.  Conditions were not photometric on the last night. Seeing during the observations generally ranged from 0".8 to 1".5.  The seeing briefly went up to 2" during the last night of observation.

The ARCONS data reduction pipeline was used to extract photon lists from the raw data files \citep{vanEyken2015}.  Each photon has an associated wavelength and timestamp.
For simplicity and to avoid any unforeseen artifacts, flat and spectral response calibration weights were not applied.  This means that the absolute measurement of flux in this analysis is not completely reliable.  For this work, what is important is the flux of the pulsar relative to the inner white dwarf, so this does not present a problem.
We used the pulsar timing software package TEMPO\footnote{http://www.atnf.csiro.au/research/pulsar/tempo.} to barycenter the photon timestamps using the JPL DE405 ephemeris and to fold them by the 2.7\,ms pulse period. Also, part of the PRESTO package was used to parse TEMPO output files \citep{presto2011}.

To predict the pulsar rotational phase for each photon, we use the basic timing model from \citet{Ransom2014}. This model constructs the Newtonian equations of motion as differential equations and solves them for the motions of the three bodies in the J0337 system. The parameters in \citet{Ransom2014} were obtained by fitting the predicted pulse arrival times to 26280 radio pulse arrival time measurements spanning 506 days, obtaining a residual scatter of 1.34\,$\mu$s. In other words, this model is capable of predicting pulse arrival times to microsecond accuracy. We updated the model by including additional observations up to 2014 October and re-fitting. 

The raw data from ARCONS are divided into five minute observation files with a gap of a few seconds between observations.  Images were made from each five minute file and were fit with a circular Gaussian point spread function (PSF) to determine an optimal aperture size for each time period.  The aperture radius was set to 1.6 times the Gaussian sigma of each PSF fit.  Due to the uncertainty in the fits, setting the aperture size as a function of time in this way was somewhat noisy, so the list of aperture radii was first smoothed with an eight-point boxcar filter. 

After this the circular apertures were applied to make a list of photons recorded within the PSF of the target object during all the observation times.  Some segments of observation time were cut, because either they were taken while the seeing had inflated to 2", the object had drifted onto a section of dead pixels, or the object had been suddenly moved during observation.  While observing, we view real time images from the science camera with one second integration times.  Sometimes the target object was difficult to see in these images.  Also, at the Coud\'{e} focus, the field would slowly rotate over time.  Periodically, we would move the telescope by a few arcseconds and then back to ensure that the target was on the intended part of the array. With the high time resolution of ARCONS these sudden moves could be accounted for by tracking the PSF as it moved and adjusting the aperture position accordingly, but the amount of data affected by this was small enough that it was simpler to discard it.  The portions of data with worse seeing were discarded because as the PSF is spread over more pixels the ratio of signal from the neutron star to background sky counts decreases.

\section{Results}

\begin{figure}
	\includegraphics[width=\columnwidth]{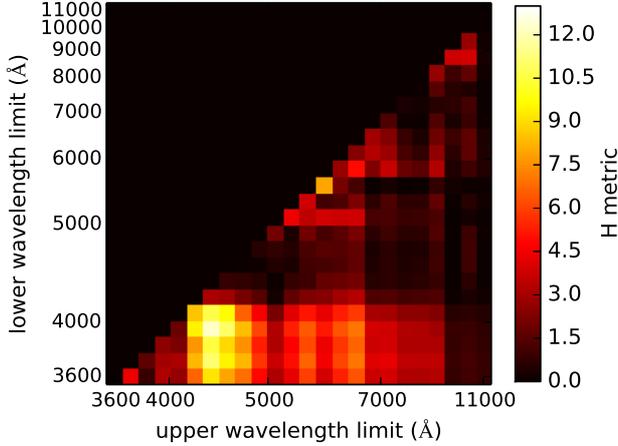}
    \caption{A search for periodicty with many wavelength cuts. The color axis gives the metrics produced by applying the H test to the photon timestamps with various wavelength cuts.  The narrow wavelength range 3950 to 4350 \angstrom\ had the highest H metric value of those tested (indicating a higher probability of periodicity), but the value obtained was not statistically significant.}
    \label{fig:wvlsearch}
\end{figure}

\begin{figure}
	\includegraphics[width=\columnwidth]{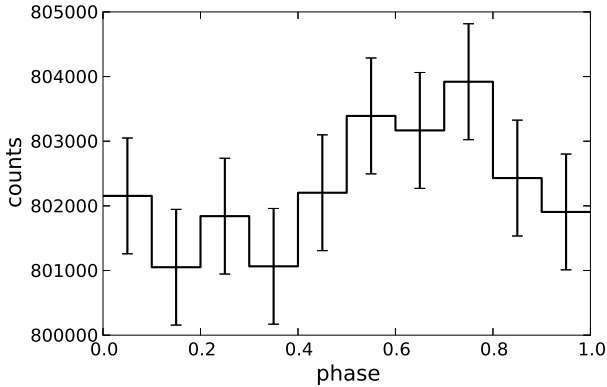}
    \caption{The observed lightcurve of J0337 as a function of phase in the band 4000-5500 \angstrom, after folding by the 2.7\,ms pulse period. The error bars for each bin are the square root of the total number of counts in the bin.  The H test shows no statistically significant detection of periodicity.}
    \label{fig:lightcurve}
\end{figure}

Once the list of photons with corresponding wavelengths and timestamps was compiled, statistical tests could be applied to test for periodicity under a given wavelength cut.
ARCONS has a broad wavelength sensitivity of 4000 to 11000 \angstrom, but the SNR in the infrared is likely lower, so some wavelength cut was needed to optimize our search for pulsed emission.  We tried to remove the arbitrariness of cutting at a particular wavelength range by testing for periodicity in many ranges, using the H test \citep{htest}.  In this test higher H values indicate more significant periodicity.  The results of the wavelength search are shown in Figure \ref{fig:wvlsearch}. The highest H metric had a value of 12.3 for the narrow wavelength range 3950 to 4350 \angstrom.  To find the significance of this H value, we created 1000 simulated photon lists with the same wavelengths as the real list, but with random pulse phases.  We then applied the H test with the same set of wavelength cuts.  We found that for 40.2\% of the simulated lists, there was one or more wavelength ranges that generated an H metric at least as high as 12.3. This implies that finding an H value of 12.3\ in the wavelength search had a low significance of 0.84 $\sigma$.

For the purpose of calculating an upper limit to the pulsed emission, we chose to use the wavelength range of 4000 to 5500\ \angstrom\ to approximately match the band pass of the Sloan Digital Sky Survey (SDSS) $g$ filter \citep{Doi2010}.  This allows for a rough comparison to the SDSS measured magnitude for the inner white dwarf.  The binned, folded light curve for the three nights of observation in the wavelength range 4000 to 5500 \angstrom\ is shown in Figure \ref{fig:lightcurve}. The error bars are the square root of the total counts in the bin.  We applied a $\chi^{2}$ test to this light curve to test for deviation from a flat line with a value set to the average count rate. The flat line would be the expected light curve if there is no pulsed emission.  The result of the test was a $\chi^{2}$ value of 10.0 with 9 degrees of freedom, corresponding to a false positive probability (FPP) of 35\%.  This indicates that the significance of pulsed emission, for the choice of binning used here, is only 0.94 $\sigma$.

For a bin-free test for periodicity we applied the H-test to the folded photon timestamps. The included photons numbered about eight million.  The H test returned an H metric of 6.04, which corresponds to an FPP of 9.1\% (or equivalently a 1.7 $\sigma$ significance). This is higher than the binned test, but it still shows no statistically significant detection of pulsed emission.

The H test provides the number of Fourier modes that maximizes the H metric, so that a Fourier model of a periodic signal can be constructed with these modes.  For our light curve, the H test indicated that the first mode (i.e. a sinusoid at the fundamental) was the most significant. We used simulated photon lists with pulse profiles of the form of our Fourier model to estimate a 95\% confidence upper limit on the pulsed optical emission. We found that in order to have a 95\% probability of observing a light curve with an H metric of 6.04 or higher, the pulsed emission (averaged over phase) would have to be at least 6.8 magnitudes fainter than the inner white dwarf companion.  Since SDSS measures the $g$ magnitude of the system to be 17.93 $\pm$ 0.01 \citep{sdss12}, this sets the 95\% confidence upper limit on the brightness of the pulsed emission at $g\sim 25$\,mag.

\section{Discussion}

Although we found no statistically significant periodicity, the lightcurve in Fig \ref{fig:lightcurve} may show a hint of periodicity that could be uncovered with more observing time.  It would be worthwhile to observe this object again with ARCONS or with one of its MKID successors that are currently in development (Meeker \etal 2016, Proc. AO4ELT, submitted).

Until there is an example of detected optical pulses in an MSP, we have no firm basis to predict how bright pulses would be. One generally useful value to predict the brightness of a pulsar is the observability metric, $\frac{\dot{E}}{d^{2}}$, where $\dot{E}$ is the rate of change in rotational energy and $d$ is the distance to the pulsar.  This metric indicates the energy available for emission in all wavelengths from a pulsar's slow down, with an assumption of isotropic emission. \citet{Thompson2000} showed that the pulsars known at the time with the highest observability metric were the ones with known high energy emission (X-ray and $\gamma$-rays). For J0337, $\frac{\dot{E}}{d^{2}} = 2.0\e{34} ergs\ s^{-1} kpc^{-2}$ \citep{Ransom2014}. This places J0337 among the pulsars with higher values, of which many are detected in X-rays (including J0337; Spiewak \etal 2016, ApJ, accepted).

\citet{Shearer2001} showed that the best predictor of optical brightness for the known optical pulsars is the strength of the B field at the pulsar's light cylinder, as formulated by \citet{Pacini1971}, with revisions in \citet{Pacini1983} and \citet{Pacini1987}.  \citet{Shearer2001} did a regression fit to find that for the five pulsars with detected optical pulses, the optical luminosity scales roughly with $\dot{E}^{1.6}$.  If this scaling holds true for MSPs, then for J0337 we would expect $m_V \sim 31$\,mag.  If this prediction is right, we will need a 30-m class telescope to detect optical emission.

If optical pulsation could eventually be found in an MSP with enough SNR, optical observations could be used for pulsar timing.  \citet{Zampieri2014} demonstrate how to compute a timing solution for the Crab pulsar using optical timing data.  Optical timing would complement ongoing radio pulsar timing arrays, because it does not suffer from interstellar dispersion or scintillation and it may be able to help characterize the sources of noise in radio timing.

\section{Conclusions}

We have presented a search for optical pulsations in the millisecond pulsar J0337+1715 at the pulse period seen in radio observations.  No significant pulsations were found in the ARCONS passband 4000-11000 \angstrom.  In the range 4000-5500 \angstrom, we have established a 95\% confidence upper limit for phase-averaged pulsed emission at 6.8 magnitudes fainter than the inner white dwarf, which sets the limit at $g \sim 25$\,mag.

\section*{Acknowledgements}

Observations were obtained with the Hale Telescope at the Palomar Observatory.  This work was supported by NSF grant AST-1411613.  Funding for the development of the MKID detectors used in this work was provided by NASA grant NNX11AD55G.  PS and SRM were supported by NASA Space Technology Research Fellowships. Fermilab is operated by Fermi Research Alliance, LLC under Contract No. De-AC02-07CH11359 with the United States Department of Energy.
The authors extend their thanks to Shri Kulkarni, Director of the Caltech Optical Observatories, and Tom Prince.

%%%%%%%%%%%%%%%%%%%%%%%%%%%%%%%%%%%%%%%%%%%%%%%%%%



\bibliographystyle{mnras}
\bibliography{mazinlab,j0337WithArcons,pulsar,msp}

%%%%%%%%%%%%%%%%%%%%%%%%%%%%%%%%%%%%%%%%%%%%%%%%%%

%%%%%%%%%%%%%%%%% APPENDICES %%%%%%%%%%%%%%%%%%%%%

%%%%%%%%%%%%%%%%%%%%%%%%%%%%%%%%%%%%%%%%%%%%%%%%%%

% Don't change these lines
\bsp	% typesetting comment
\label{lastpage}
\end{document}